\documentstyle[aps,amssymb,twocolumn,psfig]{revtex}

\def\doteqdot{=}

\begin{document}
\flushbottom
\title{Efficient bipartite quantum state purification
in arbitrary dimensional Hilbert spaces}
\author{Gernot Alber$^1$, Aldo Delgado$^1$, Nicolas Gisin$^2$, 
Igor Jex$^{1,3}$}
\address{
$^1$ Abteilung f\"ur Quantenphysik, Universit\"at Ulm,
D--89069 Ulm, Germany\\
$^2$ Group of Applied Physics, University of Geneva, 1211 Geneva 4, Switzerland\\
$^3$ Department of Physics, FJFI \v CVUT,
B\v rehov\'a 7, 115 19 Praha 1 - Star\'e M\v{e}sto, Czech Republic}
\date{\today}
\maketitle
\begin{abstract}
A new purification scheme is proposed which applies
to arbitrary dimensional bipartite quantum systems. It is based on
the repeated application of a special class of nonlinear quantum maps
and a single, local unitary operation. 
This special class of nonlinear quantum maps 
is generated in a natural way by a hermitian
generalized XOR-gate.
The proposed purification scheme offers two major advantages, namely it does
not require local depolarization operations at each step of the purification
procedure and it purifies more
efficiently than other know purification schemes.
\end{abstract}
{PACS numbers: 03.67.-a,03.67.Hk,03.65.Bz}
\section{Introduction}

Quantum information processing was proven to be superior to classical information
processing in several respects. The essence of quantum information is its ability to
employ the linearity of quantum mechanics on composite systems for practical purposes.
Many of the already demonstrated procedures \cite{Anton,2} rely on the use of highly entangled
quantum states. Entangled states are never generated in ideal form. 
Typically, either the source producing entangled quantum states or the communication channel
with which entanglement is transfered to remote
parties adds noise. 
In order to be able to exploit entanglement efficiently it
is desirable to remove as much of this additional noise as possible. This goal
can be achieved by purifying or concentrating entanglement.

The manipulation of quantum states is realized using quantum networks.  
Usually, they 
are constructed out of simple elements, so called quantum gates. 
Among them the two-particle quantum 
XOR-gate \cite{Monroe} plays a fundamental
role. In this 2-qubit gate, the first qubit controls the target qubit: if the control is in
state $|0\rangle$, the target is left unchanged, but if the control qubit is in state $|1\rangle$
the target's basis states are flipped. 
Together with one-qubit operations this gate forms a universal set of quantum
gates allowing the implementation of arbitrary unitary operations acting on qubits
\cite{gates}. 
It has been demonstrated that the quantum XOR-gate can be used for many practical tasks of
quantum information processing, such as
quantum state swapping \cite{swapping},
entangling quantum states \cite{controled swapping},
performing Bell measurements \cite{bell measurement}, dense coding \cite{dense},
and teleportation \cite{tele}.
Furthermore, in combination with selective
measurements it can be used for implementing nonlinear quantum transformations
which may be used for optimal state identification and
for state purification \cite{Bennett,gisin}. 

For many practical tasks of quantum information processing it is
desirable to extend the basic notion of such a quantum XOR-operation to
higher dimensional Hilbert spaces. Indeed, most of the physical systems that have been
proposed to hold qubits, such as
multilevel atoms or ions \cite{multilevel}
and multipath-interferometers \cite{Brendel99}, 
could equally well encode larger alphabets. 
However, there is a considerable degree of freedom involved in such a generalization. 

In the present paper we use a hermitian generalization of 
the quantum XOR-gate which applies to arbitrary dimensional Hilbert spaces and which
allows to implement a special class of nonlinear quantum transformations in a natural way.
These nonlinear transformations can
be used for the preparation of quantum states
and for efficient quantum state purification.
This will be exemplified
by discussing state purification of generalized Bell-states. These latter quantum states 
are of considerable interest in quantum information processing in higher
dimensional Hilbert spaces. 
Compared with the other known purification scheme which is valid in arbitrary dimensional
Hilbert spaces and which has been proposed by Horodecki et al. \cite{Horodecki96}
our proposed purification procedure
offers two advantages. Firstly, it does not involve a depolarization operation at
each step of the iteration procedure. Typically, it is not easy to implement 
such a depolarization operation. Furthermore,
such a depolarization operation requires a minimal size of
the ensemble of quantum states it is applied to.
Secondly, it will be demonstrated that our newly proposed method is more efficient.

The paper is organized as follows.
In Sec. II the hermitian generalized
XOR-gate is introduced which allows to implement the special class of nonlinear quantum
transformations needed for our proposed purification scheme. 
The resulting class of nonlinear quantum maps is discussed in Sec. III.
The new quantum state purification scheme, its basic properties and its efficiency
are exemplified in Sec. IV.

\section{A hermitian generalized GXOR-gate}

Let us start by summarizing characteristic properties of the
XOR-gate as they are known for qubit systems.
For qubits the action of the quantum XOR-gate 
onto a chosen set of basis
states $\left\{ \left| i\right\rangle \right\}$ with $i\in \{0,1\}$
of the Hilbert space
of each qubit is defined by
\begin{equation}
XOR_{12}\left| i\right\rangle _{1}\left| j\right\rangle _{2}\doteqdot \left|
i\right\rangle _{1}\left| i\oplus j\right\rangle_{2}~.
\label{quantum xor-gate}
\end{equation}
This transformation has the
following characteristic properties:
(i) it is unitary and thus reversible,
(ii) it is
hermitian and
(iii) $i\oplus j=0$ if and only if $i=j$. 
The first (second) index denotes the state of the control (target) qubit 
and $\oplus $ denotes addition modulo(2).

Let us now consider the problem of generalizing the quantum XOR-gate to
higher dimensional Hilbert spaces.
The desired generalized quantum XOR-gate (GXOR-gate) should
act on two $D$ -dimensional quantum systems.
In analogy with qubits we will call these two systems 
qudits.
The basis states $|i\rangle $ of each qudit are labeled by elements in
the ring ${\sf Z}_{D}$ which we denote by the numbers $i=0,...,D-1$ with the
usual rules for addition and multiplication {\it modulo(D)}. In
principle, the GXOR-gate could be defined in a straightforward way
by using 
Eq.(\ref{quantum xor-gate}) and by performing $i\oplus j$ 
{\it modulo(D)}, i. e.
\begin{equation}
GXOR_{12}\left| i\right\rangle _{1}\left| j\right\rangle _{2}\doteqdot \left|
i\right\rangle _{1}\left| i\oplus j\right\rangle _{2}.
\label{error}
\end{equation}
The GXOR-gate defined in 
(\ref{error}) is unitary but not hermitian for $D>2$. Therefore it is no longer
its own inverse. 
Thus, the inverse GXOR-gate has to be obtained from the 
GXOR-gate of Eq.(\ref{error}) by iteration, i.e.
$GXOR_{12}^{-1}=(GXOR_{12})^{D-1}=GXOR_{12}^{\dagger}\neq GXOR_{12}$.
All these inconvenient properties of this preliminary definition (\ref{error})
can be removed by the alternative definition
\begin{equation}
GXOR_{12}\left| i\right\rangle_{1}\left| j\right\rangle _{2}\doteqdot \left|
i\right\rangle _{1}\left| i\ominus j\right\rangle _{2}.
\label{generalized xor-gate}
\end{equation}
In Eq.(\ref{generalized xor-gate}) $i\ominus j$ denotes the difference
$i-j$ modulo (D). In the special case of qubits the definition of
Eq.(\ref{generalized xor-gate}) reduces to Eq.(\ref {quantum xor-gate})
as $i\ominus j\equiv i\oplus j~modulo(2)$. Furthermore, this 
definition preserves all the properties of Eq.(\ref{quantum xor-gate}) 
also for arbitrary values of $D$, namely it is unitary,
hermitian and $i\ominus j=0 ~modulo(D)$ if and only if $i=j$. 

The GXOR-gate of Eq.(\ref{generalized xor-gate})
admits a natural extension to control and target systems 
with continuous spectra. In this case the basis states $|i\rangle$
are replaced
by the basis states $\left\{ \left| x\right\rangle \right\}$ with
the continuous variable $x \in {\bf R}$.
These new basis states are assumed to satisfy the 
orthogonality condition $\left\langle x \right| y \rangle=
\delta\left( x-y \right)$.
Furthermore, as the dimension $D$ tends to infinity the modulo
operation entering Eq.(\ref{generalized xor-gate}) can be omitted.
Thus, for continuous variables the action of the GXOR-gate becomes
\begin{equation}
GXOR_{12}\left| x\right\rangle_{1}\left| y\right\rangle_{2}=
\left| x\right\rangle_{1}\left| x-y\right\rangle_{2}.
\label{continuous xor-gate}
\end{equation}
Let us note that this definition for the case of continuous variables
is different from the 
generalized XOR-gate proposed in 
Ref. \cite{braunstein}. This latter gate is not
hermitian whereas the GXOR-gate of
Eq.(\ref{continuous xor-gate}) is both unitary and hermitian.
The GXOR-gate of Eq. 
(\ref{continuous xor-gate}) can be represented in terms of
a translation 
and a space inversion, namely
\begin{equation}
GXOR_{12}\left| x\right\rangle_{1}\left| y\right\rangle_{2}=
\hat{\Pi}_{2}
{\it e}^{(-i\hat{P}^{(2)}_{y}\hat{x}^{(1)})}
\left| x\right\rangle_{1}
\left| y\right\rangle_{2}.
\end{equation}
Thereby $\hat{P}^{(2)}_{y}$ denotes the canonical momentum operator
which is conjugate to the
position operator $\hat{y}^{(2)}$ acting on quantum system 2 
and 
$\hat{\Pi}_2$ is the corresponding operator of space inversion.

Let us discuss a possible
physical realization of the 
GXOR-gate defined by Eq.(\ref{generalized xor-gate}) 
which is based on
nonlinear optical elements.
For this purpose we assume that the two quantum systems which are going to be
entangled are two modes of the radiation field. The basis states $|i\rangle_1$ $(i=0,...,D-1)$
of the first quantum system are formed by $n$-photon states 
of  mode one with $0\leq n\leq D-1$.
The basis states of the second quantum system $|k\rangle_2$
$(k=0,...,D-1)$ are formed by
Fourier transformed $n$-photon states of this latter mode,
i.e. $|k\rangle_2 = 1/\sqrt{D}\sum_{n=0}^{D-1}
{\rm exp}(i2\pi kn/D)|n\rangle_2$.
Let us further assume that the dynamics of these two modes of the
electromagnetic field are governed by the Kerr-effect \cite{Kerr}. Thus, in the interaction 
picture their Hamiltonian  is given by
$H=\hbar\chi a_1^{\dagger}a_1 a_2^{\dagger}a_2$ with the creation and annihilation
operators $a_{1,2}^{\dagger}$ and $a_{1,2}$ of modes $1$ and $2$, respectively.
For the sake of simplicity 
the nonlinear susceptibility $\chi$ is assumed to be real-valued and positive.
Preparing intitially both quantum systems in state
$|i\rangle_1|k\rangle_2$ after an interaction time
of magnitude $t=2\pi/(D\chi)$ 
this two-mode system ends up in state
$|\psi\rangle_{12} = |i\rangle_1|k-i\rangle_2$. Applying to this latter state
a time reversal transformation which may be implemented by the process of
phase conjugation \cite{Kerr}
we finally arrive at the desired state $|i\rangle_1|i-k\rangle_2$.
Thus this combination of a Kerr-interaction with a time reversal transformation
is capable of realizing the GXOR-gate of Eq. (\ref{generalized xor-gate}).

As the GXOR-gate of Eq.(\ref{generalized xor-gate}) differs from the alternative definition
of Eq.(\ref{error}) by local unitary operations only both are expected to
exhibit similar properties as far as entanglement operations are concerned.
However,
with the help of this GXOR-gate alone already
a variety of interesting quantum operations can be implemented without having to
use additional local unitary transformations. 
In addition, as will be shown in Secs. III and IV 
this entanglement operation is particularly useful in the context of quantum state
purification for implementing nonlinear quantum state
transformations.

As a first application let us consider the preparation of a 
basis of entangled states from  separable ones. If $|l\rangle |m\rangle$
with $l,m,=0,...,D-1$ denotes an orthonormal basis of factorized states 
an associated basis of entangled 
two-particle states is given by
\begin{eqnarray}
|\psi_{lm}\rangle = GXOR_{12}[(F|l\rangle)_1|m\rangle_2].
\label{Bell}
\end{eqnarray}
Thereby $F$ denotes the discrete Fourier transformation, i.e.
$F|l\rangle = (1/\sqrt{D})\sum\limits_{k=0}^{D-1} {\em \exp}(i 2\pi lk/D) |k\rangle$.
For qubits this unitary quantum transformation leads to the well known basis
of four Bell states.
In the simplest higher dimensional case of $D=3$,
for example, the first four states of this entangled
generalized Bell basis are given by
\begin{eqnarray}
|\psi_{00}\rangle &=&\frac{1}{\sqrt{3}}[|00\rangle + |11\rangle +|22\rangle],\nonumber\\
|\psi_{10}\rangle &=&\frac{1}{\sqrt{3}}
[|00\rangle + e^{i2\pi/3}|11\rangle +e^{-i2\pi/3}|22\rangle],\nonumber\\
|\psi_{20}\rangle &=&\frac{1}{\sqrt{3}}
[|00\rangle + e^{-i2\pi/3}|11\rangle +e^{i2\pi/3}|22\rangle],\nonumber\\
|\psi_{01}\rangle &=&\frac{1}{\sqrt{3}}[|02\rangle + |10\rangle +|21\rangle].
\end{eqnarray}
As the GXOR-gate is hermitian it can also be used to disentangle this basis
of generalized Bell states again by inverting Eq.(\ref{Bell}).
This basic disentanglement property is
of practical significance. It enables to reduce Bell
measurements to measurements of factorized states.
Examples where these latter types of
measurements are of central interest are
dense coding \cite{dense} and quantum teleportation schemes
\cite{tele}.

\begin{figure}
\centerline{\psfig{figure=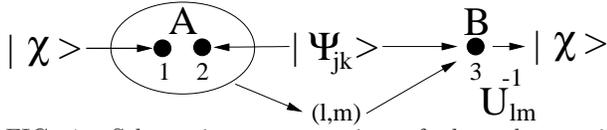,width=8.0cm,clip=}}
\caption{Schematic representation of the teleportation scheme involving 
Bell measurements onto the generalized Bell states of Eq.(\ref{Bell}).}
\label{Fig.1}
\end{figure}

The basis of entangled Bell
states resulting from Eq.(\ref{Bell}) can be used for teleporting an arbitrary
D-dimensional quantum state from A (Alice) to B (Bob).
For this purpose let us assume that A and B share an entangled pair of
particles prepared in state
$|\psi_{lm}\rangle$ as defined by Eq.(\ref{Bell}). If A wants to teleport an
unknown quantum state $|\chi\rangle = \sum\limits_{n=0}^{D-1}\alpha_n |n\rangle$ to B
she has to perform a Bell measurement which yields one of the entangled basis
states of Eq.(\ref{Bell}) as an output state (compare with Fig. (\ref{Fig.1})). 
Conditioned on the measurement result of Alice, Bob has to perform an appropriate unitary
transformation onto his particle which prepares this latter
particle in state $|\chi\rangle$.
This arbitrary dimensional teleportation scheme rests on the identity
\begin{eqnarray}
|\chi\rangle |\psi_{j k}\rangle_{23} &=&
\sum_{l, m = 0}^{D-1} |\psi_{l m}\rangle_{12}
\frac{e^{-i2\pi j m/D}}{D} U_{l m} |\chi\rangle,\nonumber\\ 
U_{l m }|n\rangle &=& 
e^{-i2\pi n(l - j)/D} |n - k - m\rangle.
\label{tele}
\end{eqnarray}
This basic relation for teleportation for an arbitrary dimensional state
$|\chi\rangle$ 
can be derived in a straightforward way from Eqs.
(\ref{generalized xor-gate}) and (\ref{Bell}). 
The classical communication requires $2\log_2(D)$ bits, which is the minimum
necessary in all quantum teleportation schemes.


\section{Nonlinear quantum maps on density matrices}

With the help of the hermitian
GXOR-gate of Eq.(\ref{generalized xor-gate}) an interesting class of  
nonlinear quantum maps can be implemented in a natural way.
Together with filtering measurements acting on a target quantum system $t$
the GXOR-gate of Eq. (\ref{generalized xor-gate}) induces
nonlinear transformations of quantum states of a control system $c$. This
can be demonstrated most easily by considering the case of two qudits which
are prepared in the quantum states $\sigma^{t}$ and $\sigma^{c}$ initially. 
Let us perform the quantum operation
\begin{eqnarray}
T(\sigma ^{c},\sigma ^{t}) &\doteqdot &\frac{A\left( \sigma ^{c}\otimes
\sigma ^{t}\right) A^{\dagger }}{Tr[A\left( \sigma ^{c}\otimes \sigma
^{t}\right) A^{\dagger }]}  
\label{non-linear transformation 1}
\end{eqnarray}
on these two qudits with
\begin{eqnarray}
A &\doteqdot &({\bf 1_{c}}\otimes P)~GXOR_{ct}. 
\label{Ain}
\end{eqnarray}
Thereby ${\bf 1_{c}}$ denotes the identity operator acting in the Hilbert
space of the control system and $P=\left|p\right\rangle_{tt}\left
\langle p\right|$ is the projector onto the state $\left| p\right\rangle_{t}$ of the 
target qudit.
With the
decomposition 
\begin{eqnarray}
\sigma^{c}&=&\sum\limits_{ij}^{D-1}\sigma_{ij}^{c} 
\left| i\right\rangle_{cc}\left \langle j\right|,\nonumber
\\
\sigma^{t}&=&\sum\limits_{ij}^{D-1}\sigma_{ij}^{t} 
\left| i\right\rangle_{tt}\left \langle j\right| .
\end{eqnarray}
Eqs. (\ref{non-linear transformation 1}) and 
(\ref{Ain}) may be rewritten in the form
\begin{equation}
T(\sigma^{c},\sigma^{t})=\frac{\sum\limits_{ijkl}^{D-1}\sigma_{ij}^{c}\sigma_{kl}^{t}
\left| i\right\rangle_{cc}\left \langle j\right|\otimes
P\left| i\ominus k\right\rangle_{tt}\left \langle j\ominus l\right|P}
{\sum\limits_{ikl}^{D-1}\sigma_{ii}^{c}\sigma_{kl}^{t}
\left\langle 0| i\ominus k\right\rangle_{tt}\left \langle i\ominus l|0\right\rangle}.
\end{equation}
Assuming that both control and target qudit are 
prepared in the same state initially, i.e. 
$\sigma^{c}\equiv \sigma^{t}$, it turns out that
Eq.(\ref{non-linear transformation 1}) is equivalent to the relations 
\begin{eqnarray}
T(\sigma^{c},\sigma^{t}& \equiv& \sigma^{c})=
\sigma_{output}^{c}\otimes P,\nonumber\\
\sigma_{output}^{c}&=&\frac{\sum\limits_{ij}^{D-1} \sigma_{i,j}^{c} \sigma^{c}_{i-p,j-p}
\left| i\right\rangle_{cc}\left \langle j\right|} {\sum\limits_{i}^{D-1}
\sigma_{ii}^{c}\sigma^{c}_{i-p,i-p}} .  
\label{final state}
\end{eqnarray}
As a result of the quantum operation (\ref{non-linear transformation 1})
the combined system formed by the control and the target qudit forms a
factorizable state with the target qudit being
in state $|p\rangle \langle p|$.
According to Eq.(\ref{final state}) the density matrix elements of $\sigma^c$
with respect to the computational basis $|i\rangle$ $(i=0,...,D-1)$ have been
multiplied with each other. 
The final state of the control qudit is prepared with probability
$p_c = \sum\limits_i^{D-1}\sigma_{i,i}^c \sigma^c_{i-p,i-p}$.
In the case of qubits, for instance, 
the nonlinear transform of the (unnormalized) control density matrix is given by
\begin{eqnarray}
\sigma^{c}_{output} = \left(
\begin{array}{cccc}
\left(\sigma^c_{00}\right)^2 & \left(\sigma^c_{01}\right)^2 & \left(\sigma^c_{02}\right)^2 & \left(\sigma^c_{03}\right)^2 \\
\left(\sigma^c_{10}\right)^2 & \left(\sigma^c_{11}\right)^2 & \left(\sigma^c_{12}\right)^2 & \left(\sigma^c_{13}\right)^2 \\
\left(\sigma^c_{20}\right)^2 & \left(\sigma^c_{21}\right)^2 & \left(\sigma^c_{22}\right)^2 & \left(\sigma^c_{23}\right)^2 \\
\left(\sigma^c_{30}\right)^2
& \left(\sigma^c_{31}\right)^2 & \left(\sigma^c_{32}\right)^2 & \left(\sigma^c_{33}\right)^2 \\
\end{array}\right) , \label{sqb0} 
\end{eqnarray}
if projected onto the state $\vert 0\rangle_t$. The elements of the original density matrix have been 
squared. If one projects onto the state $\vert 1\rangle_t$, the original density
matrix elements are mixed in a more complicated way and one obtains the (unnormalized) density
matrix
\begin{eqnarray}
\sigma^{c}_{output} = \left(
\begin{array}{cccc}
\sigma^c_{00}\sigma^c_{33} & \sigma^c_{01}\sigma^c_{32} & \sigma^c_{02}\sigma^c_{31} & 
\sigma^c_{03}\sigma^c_{30} \\
\sigma^c_{10}\sigma^c_{23} & \sigma^c_{11}\sigma^c_{22} & \sigma^c_{12}\sigma^c_{21} & 
\sigma^c_{13}\sigma^c_{20} \\
\sigma^c_{20}\sigma^c_{13} & \sigma^c_{21}\sigma^c_{12} & \sigma^c_{22}\sigma^c_{11} & 
\sigma^c_{23}\sigma^c_{10} \\
\sigma^c_{30}\sigma^c_{03} & \sigma^c_{31}\sigma^c_{02} & \sigma^c_{32}\sigma^c_{01} & 
\sigma^c_{33}\sigma^c_{00} \\
\end{array}\right) . \label{sqb1}
\end{eqnarray}

From Eq. (\ref{final state}) it is easy to verify that the quantum 
operation (\ref{non-linear transformation 1}) has the following basic
properties: (i) it maps density matrices onto density matrices, (ii) it is 
not injective and nonlinear, and (iii) there are states invariant under this 
transformation.
It is also possible to extended the quantum operation of 
Eq. (\ref{non-linear transformation 1}) to cases in which there is more than one
control system and in which both the control and the
target systems are composite quantum systems each of which consists of M
qudits. In this case $\sigma^{c}$ describes a general $M$-qudit state of
the form
\begin{equation}
\sigma^{c}=\sum_{{\bf ij}}\sigma_{{\bf ij}}^{c}\left| {\bf i}%
\right\rangle_{cc}\left \langle {\bf j}\right|,
\end{equation}
with ${\bf i}=(i_1,...,i_M)$ and ${\bf j}=(j_1,...,j_M)$. In Eq.(\ref
{non-linear transformation 1}) the operator $A$ has to be replaced by
\begin{eqnarray}
A &\doteqdot &({\bf 1_{c}}\otimes P)\Pi
_{j=1}^{M}\Pi _{i=1}^{N}GXOR_{ct_{i}}^{(j)}  
\label{finA}
\end{eqnarray}
with the projection operators $P=\prod_{i=1}^{N}\otimes
P_{t_{i}}$ and $P_{t_{i}}=\left| {\bf p_i}
\right\rangle_{t_{i}t_{i}}\left\langle {\bf p_i}\right|$ onto state $\left| 
{\bf p_i}\right\rangle_{t_{i}}$ of the $M$-qudit target system $t_i$. Thereby
the GXOR-gate $GXOR^{(j)}_{ct_i}$ operates on the $j$-th qudit of the control
and of the $i$-th target system.
For the special case of $p_i=0$ for all $t_i$, for example,
the resulting final state of the 
control system is given by
\begin{equation}
\sigma_{output}^{c}=
\frac{\sum\limits_{{\bf ij}}(\sigma_{{\bf ij}}^{c})^{1+N}\left| 
{\bf i}\right\rangle_{cc}\left \langle {\bf j}\right|}
{\sum\limits_{{\bf i}%
}(\sigma_{{\bf ii}}^{c})^{1+N}}.
\label{final register state arbitrary powers}
\end{equation}
and is prepared with probability
$p_c = \sum\limits_{{\bf i}} (\sigma^c_{{\bf ii}})^{1+N}$.


\section{Bipartite purification in higher dimensional spaces}

In general, for $N=1$ the nonlinear quantum transformation of Eq.
(\ref{final register state arbitrary powers}) 
has not only invariant states for $P_{t_1} = |0\rangle_{t_1 t_1}\langle 0|$ 
but also for other projectors $P_{t_1}$.
This suggests to use this 
nonlinear quantum transformation for the purification of quantum states of a
two-qudit system. 
For the special case of a control system consisting of two-qubits
such a purification scheme which is based on the nonlinear quantum transformation
of Eq.(\ref{final register state arbitrary powers})
has already been
proposed previously \cite{gisin}. 
In order to discuss an analogous purification scheme in arbitrary dimensional
Hilbert spaces
we start from the observation that for $M=2$ the entangled basis
state $|\psi_{00}\rangle$ of Eq.(\ref{Bell}) is a fixed point of the nonlinear
two-particle quantum map of Eq.(\ref{final register state arbitrary powers}).
Inspection of the two `squared' matrices (\ref{sqb0}) and (\ref{sqb1}) 
shows that this is true not only for the projector
$P_{t_1}=\vert 0\rangle_{tt}\langle 0\vert$ but also for 
$P_{t_1}=\vert 1\rangle_{tt}\langle 1\vert$. 
Therefore, this nonlinear quantum transformation may be used for purifying
quantum states towards the entangled state $|\psi_{00}\rangle$.
Thereby the possibility to use
both projectors in the purification process
increases its efficiency considerably. 

Purification in higher dimensional Hilbert
spaces has been considered previously by Horodecki et al. \cite{Horodecki96}. These 
authors generalized the approach of Bennett et al. \cite{Bennett} to
arbitrary dimensional Hilbert spaces. The purpose of their protocol
is to distill Bell states from a noisy channel. Their protocol combines two
basic steps, namely a nonlinear quantum map which 'squares' the density matrix
elements and a depolarizing channel  converting the resulting output state
into a Werner state. The depolarizing channel guarantees that at each step of
the purification protocol an initially prepared Werner state is mapped again
onto a Werner state with a higher admixture of the Bell state. This protocol
is capable of purifing all non-separable Werner states in arbitrary
dimensional Hilbert spaces. The depolarization operation involved in
this purification protocol requires that suitably chosen local unitary
operations have to be applied to a sufficiently large number of two-qudit
systems. For the case of qubits it has been shown that such a depolarization
may be achieved with a set of 12 suitably chosen unitary operations
\cite{Bennettdepol}. For qudits with $D>2$ suitable minimal numbers of
unitary operations  are not known at present. 
In view of these inconveniences in implementing a depolarizing operation it
appears desirable to develop alternative purification strategies which do not
involve such a depolarization procedure.  For the case of qubits such a
procedure has already been developed by Deutsch et al. \cite{Deutsch}. In the
following we propose such a method which applies to arbitrary dimensional
Hilbert spaces. It is based on the nonlinear quantum transformation
of Eq.(\ref{final register state arbitrary powers}) (with $N=1$)
followed by a single local unitary operation acting on both qudits. Thus,
this proposed purification procedure does not require a minimal number of
two-qudit systems at each step of the purification protocol. 

In order to exemplify basic properties of our purification scheme let
us consider the purification of a Werner state of the form

\begin{equation}
\sigma^{c} = \lambda |\psi_{00}\rangle \langle \psi_{00}|
+ (1-\lambda){\bf 1}/D^2.
\label{Werner}
\end{equation}

\noindent where the parameter $\lambda$ is related to the fidelity
$F=\langle\psi_{00}|\sigma^c|\psi_{00}\rangle$ through the expression

\begin{equation}
F=\lambda+\frac{1-\lambda}{D^2}.
\label{fidelity}
\end{equation}

\noindent The state (\ref{Werner}) may result from a physical situation where
two spatially separated parties, say A(lice) and B(ob), want to share the
entangled basis state $|\psi_{00}\rangle$ but with a probability of
$(1-\lambda)$ the transmission of this entangled pair through a quantum
channel leads to unwanted noise represented by the chaotic state ${\bf
1}/D^2$. This initial quantum state $\sigma^{c}$ is non-separable if and only
if $\lambda >\lambda_D = (1 + D)^{-1}$ \cite{Ruben} so that a purification
scheme  can succeed only for those values of $\lambda$. 

We propose a purification scheme which is based on a two-step iteration
procedure. In the first part of each iteration step we apply the nonlinear
quantum map of Eq.(\ref{final register state arbitrary powers}) (with $N=1$)
by projecting onto an arbitrary two-qudit target state $|ii\rangle \langle
ii|$  ($i=0,...,D-1$). Correspondingly, the initially prepared
Werner state (\ref{Werner}) is converted into the quantum state

\begin{equation}
\sigma^{c(1)}_{output} = \lambda_1 |\psi_{00}\rangle \langle \psi_{00}|
+ \lambda_2 {\bf 1}/D^2 + \lambda_3 \sum_{i=0}^{D-1} \vert ii\rangle\langle ii\vert
\label{outputint}
\end{equation}

\noindent with the coefficients $\lambda_i$ depending on the initial choice
of $\lambda$. However, it turns out that both the pure state
$|\psi_{00}\rangle \langle \psi_{00}|$ and the mixed state
$\sum\limits_{i=0}^{D-1}|ii\rangle \langle ii|$ are fixed points of the
nonlinear quantum transformation of Eq.(\ref{final register state arbitrary
powers}). Therefore, an additional local unitary transformation is required in
order to guarantee convergence of the iteration procedure towards the desired
final state $|\psi_{00}\rangle \langle \psi_{00}|$.  Thus, in the second part
of each iteration step parties A and B perform a local twirling operation
\cite{Horodecki96} $U_A\otimes U_B^*$, i.e.

\begin{equation}
\sigma^{c(1)}_{output} \to
\sigma_{output}^{c(2)}\equiv
U_A\otimes U_B^*\sigma^{c(1)}_{output} U_A^{\dagger}\otimes U_B^{*\dagger}.
\label{pur}
\end{equation}

\noindent Ultimately this transformation achieves convergence towards our
desired final state $|\psi_{00}\rangle \langle \psi_{00}|$ by altering the
mixed state $\sum_{i=0}^{D-1}|ii\rangle \langle ii|$ but still leaving state
$|\psi_{00}\rangle \langle \psi_{00}|$ invariant. Therefore, the depolarization
operation of the protocol of Horodecki et al. \cite{Horodecki96} is replaced
by a single twirling operation. For the unitary transformation involved in
this twirling operation we propose to choose a discrete Fourier transform, i.e.

\begin{equation}
U_{A(B)}|k\rangle_{A(B)} =
\frac{1}{\sqrt{D}} \sum_{n=0}^{D-1} {\rm exp}^{(i2\pi k n/D)}|n\rangle _{A(B)}.
\label{Fourier}
\end{equation}

\noindent This choice is motivated by the desire to increase the success
probability of the purification process and to maximize the radius of
convergence of the iteration procedure. As both the intermediate output state
of Eq.(\ref{outputint}) and the unitary transformation of Eq.(\ref{Fourier})
are invariant under transformations of the basis states of the form $|i\rangle
\to |i+1\rangle$ ($i=0,...,D-1$), all the projections $P_{t_1} = |ii\rangle
\langle ii|$ yield the same success probability thus increasing
efficiency. Iterating the two-step procedure based on the nonlinear quantum
transformation of Eq.(\ref{final register state arbitrary powers})
and the local twirling operation of Eq.(\ref{pur}) yields our proposed
purification procedure.

Numerical simulations performed for dimensions $2\leq D \leq 20$ demonstrate 
that this purification procedure is capable of purifying almost all non-separable
Werner states of the form of Eq.(\ref{Werner}). The dependence of 
its range of convergence on the dimension $D$ of the Hilbert space is
apparent from Fig.3. The dashed line indicates the minimum values 
$F_{c} = 1/D$ of the fidelity for which the Werner state of Eq.(\ref{Werner})
is still non-separable. The purification protocol of Horodecki et al.
\cite{Horodecki96} converges for all initial values $F>F_c$. The solid line
represents the minimal initial value of the fidelity for which our protocol
purifies. It is clear from Fig.3 that the range of convergence of our
purification scheme is slightly smaller but approaches the ideal limit $F_c$
with increasing dimension $D$ of the Hilbert space. Already for moderately
large dimensions $D$ our  range of convergence approaches the ideal range
closely.

Let us now compare the efficiency of our purification protocol with the one
proposed by Horodecki et al. \cite{Horodecki96}. In particular, we are
interested in answering the question, how many iterations are needed to
obtain state $|\psi_{00}\rangle$ with a prescribed final fidelity
$F_{final}=\langle \psi_{00}| \sigma^{c}_{final}|\psi_{00}\rangle$ for a given
dimension $D$? In order to clarify the calculation of this efficiency let us
briefly reconsider the basic steps involved in a purification protocol. They
are represented  schematically in Fig. 4. Initially the purification process
starts with an ensemble of $N_{initial}$ Werner states each of which is
described by Eq.(\ref{Werner}). In each step of the iteration procedure the
ensemble of two-qudit states is divided into two equal parts which serve as
control and target systems. The nonlinear quantum transformation is performed
by projecting onto one of the target states $P_{t_1}=|ii\rangle \langle ii|$
($i=0,...,D-1$). As our initial state and our unitary transformation of
Eq.(\ref{Fourier}) are invariant under the transformation $|i\rangle \to
|i+1\rangle$ of the basis states, all these projections are equally probable
despite the fact that our purification procedure does not yield a Werner state
at each step of the iteration procedure. This nonlinear quantum transformation
is followed by a local twirling transformation (compare with Eq.(\ref{pur}))
which is based on the discrete Fourier transformation of Eq.(\ref{Fourier}).
Thus, after the first step of our purification procedure we are left with
$[p_1\times N_{initial}/2]$ purified two-qudit systems. Thereby $p_1$ denotes
the probability of obtaining the target qudit in one of its basis states
$|i\rangle$ ($i=0,...,D-1$).  Continuing this iteration procedure after $n$
iterations the number of remaining purified two-qudit systems is given by

\begin{equation}
N_{purified} = \frac{N_{initial}}{2^n}\prod\limits_{l=1}^{n}p_l.
\label{eta}
\end{equation}

\noindent 
Accordingly, the efficiency $\eta$ of this purification process is
given by $\eta = N_{purified}/N_{initial}$.

In Fig.5 and Fig.6
the dependence of the efficiency $\eta$ on the
initial fidelity of the Werner state
$F$ (compare with Eq.(\ref{fidelity}))
is depicted for dimensions $D=6$ and $D=9$ and for different values of the final fidelity
$F_{final} = \langle \psi_{00}|\sigma_{final}^c |\psi_{00}\rangle$. From
these figures it is apparent that  our protocol requires
fewer steps than the protocol of Horodecki et al. \cite{Horodecki96}. Furthermore,
numerical studies also indicate that in both purification protocols the success probabilities
$p_l$ entering Eq.(\ref{eta}) are comparable in magnitude. Thus, the overall better
efficiency of our purification protocol which is apparent from Figs. 5 and 6
reflects the fewer number of steps $n$ which
are required for achieving a given final accuracy. With increasing
accuracy of the final purified
state this difference in efficiencies between both purification protocols
becomes larger and larger. From Figs. 5 and 6 one also notices a second characteristic feature which has been
found also in other numerical simulations. For a given value of the final fidelity
the differences between the efficiencies of both protocols becomes smaller with increasing
dimension $D$ of the Hilbert space involved.

\section{Conclusions}

A novel purification scheme has been proposed.
It is based on the iterative application of
a special class of nonlinear quantum maps and a single, local unitary transformation.
The required nonlinear quantum map can be implemented conveniently by a hermitian 
generalized quantum XOR-gate.
The proposed purification scheme has several attractive features. 
Firstly, it applies to arbitrary dimensional bipartite quantum systems.
Secondly, it does not require a depolarization operation at each
step of the iteration procedure. It rests on a single twirling operation which
is performed at each step of the iterative purification scheme.
Thirdly, the proposed procedure achieves purification in a very efficient
way. In particular, it has been demonstrated 
that it achieves purification of
Werner states in a more efficient way than the other known purification protocol
which has been introduced by Horodecki et al. \cite{Horodecki96}.
Furthermore, its almost maximal range of convergence 
indicates
that the employed local twirling operation which is based on a
discrete Fourier transform is a good choice.
Such a transformation can be implemented easily in many 
quantum systems.

The proposed purification method
may also be generalized to multi-partite quantum systems.
In this context it would be particularly interesting to develop efficient purification
protocols for GHZ-like quantum states.

This work is supported by
the DFG (SPP `Quanteninformationsverarbeitung'),
by the ESF program on
`Quantum Information Theory and Quantum Computation', by the
European IST-1999-11053 EQUIP project, and by the DLR (CZE00/023).
A.D. acknowledges support by the DAAD.
I.J is supported  
by the Ministry of Education and the GA of the Czech Republic. 
We are grateful to Steve Barnett for stimulating discussions.

\newpage

\begin{figure}
\centerline{\psfig{figure=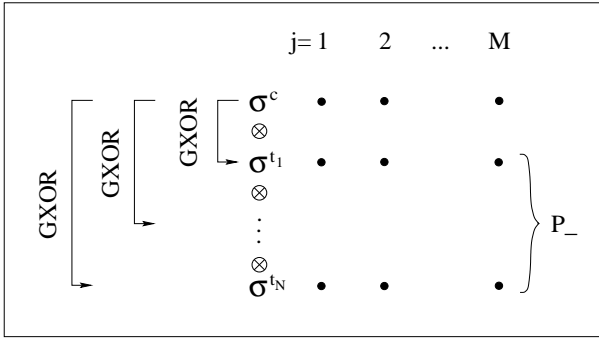,width=8.0cm,clip=}}
\caption{Schematic representation of the GXOR-gates and projections
involved in the nonlinear quantum transformation 
of Eq. (\ref{final register state arbitrary powers}). The qudits are represented
by dots. The dots of the first line represent the $M$ qudits of the control
system. The dots of the following lines represent the $M\times N$ qudits of the $N$
target systems $t_{1},t_{2},...,t_{N}$. The GXOR-gate 
$GXOR_{ct_i}^{(j)}$
acts on the $j-th$ qudit
of the control and target system $t_{i}$ with
$j\in\left\{1,2,...,M\right\}$ and $i\in\left\{1,2,...,N\right\}$. The 
operator $P_{-}$ projects the state of the whole
systems onto state $\left| {\bf 0}\right\rangle \left\langle {\bf 0}\right|$ 
with $\left| {\bf 0}\right\rangle=\left| 0\right\rangle_{1}
\left| 0\right\rangle_{2}...\left| 0\right\rangle_{MN}$.}
\label{Fig.2}
\end{figure}

\begin{figure}
\centerline{\psfig{figure=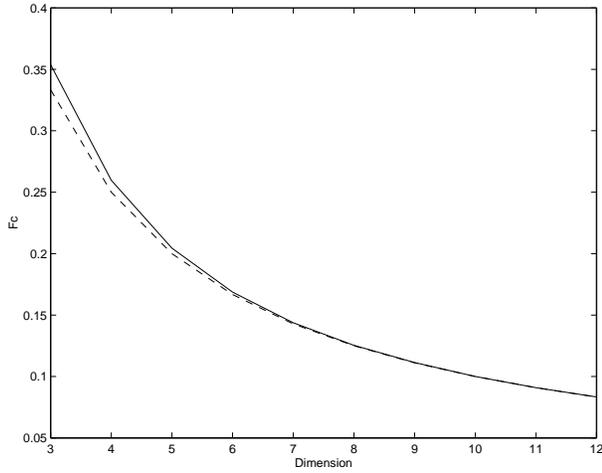,width=8.0cm,clip=}}
\caption{Dependence of the minimal initial fidelity
$F_{c}$ needed to purify a Werner state (compare with Eq. (\ref{Werner}))
as a function of the
dimension $D$: our protocol (full line), Horodecki's protocol (dashed line).}
\label{Fig.3} 
\end{figure}

\begin{figure}
\centerline{\psfig{figure=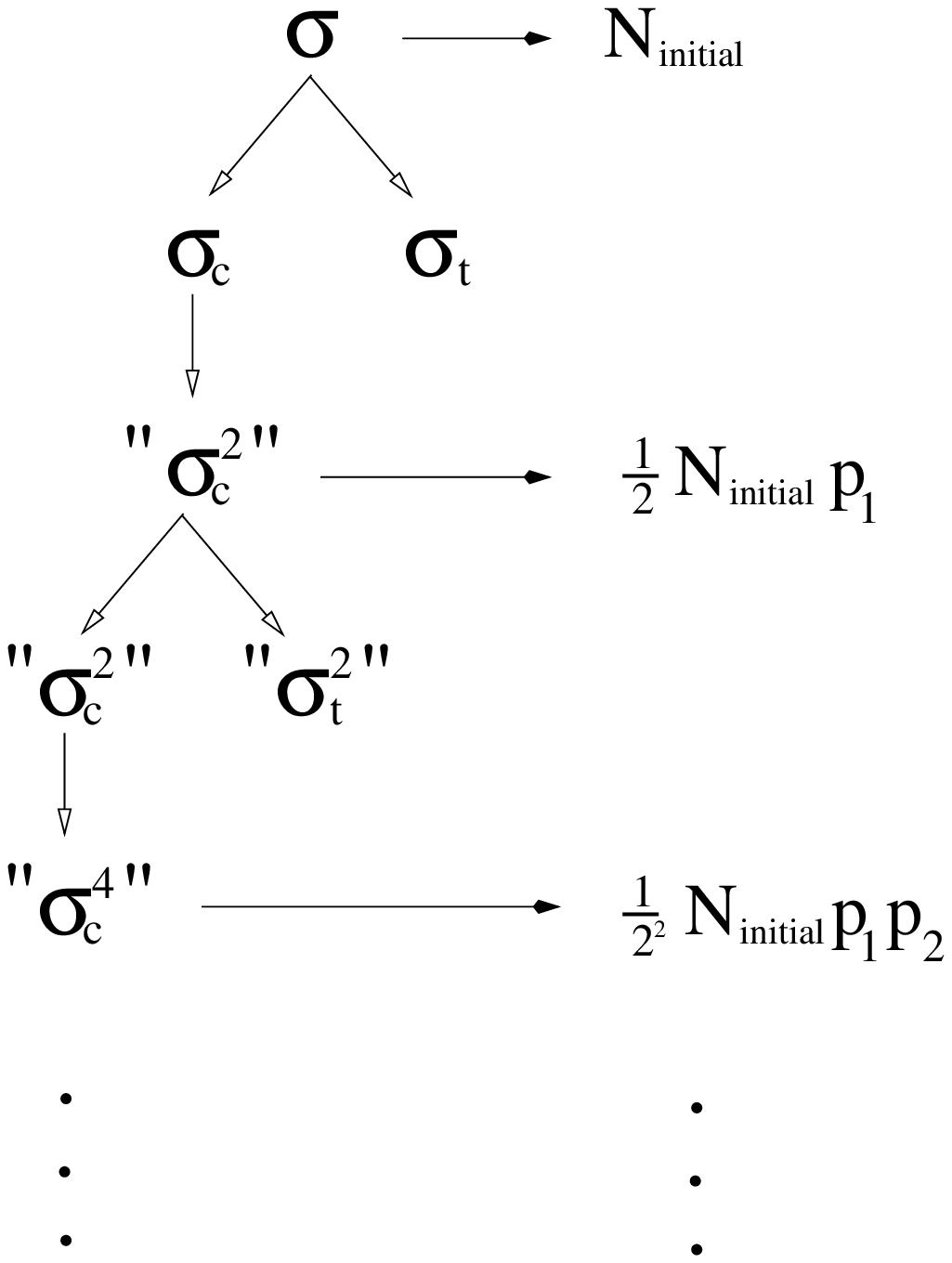,width=8.0cm,clip=}}
\caption{This scheme illustrates the calculation of the efficiency $\eta$. 
The initial collection of qudits is split into two equal parts. One
part forms the control, the other the target qudits. The success probability of the
process after the first step is $p_1$, the total number of `squared' qudits is
given by $(1/2)N_{initial} p_1$. 
The procedure is repeated until the required fidelity $F_{final}$ for the resulting state 
is reached after $n$ steps.}
\label{Fig.6}
\end{figure}
\begin{figure}
\centerline{\psfig{figure=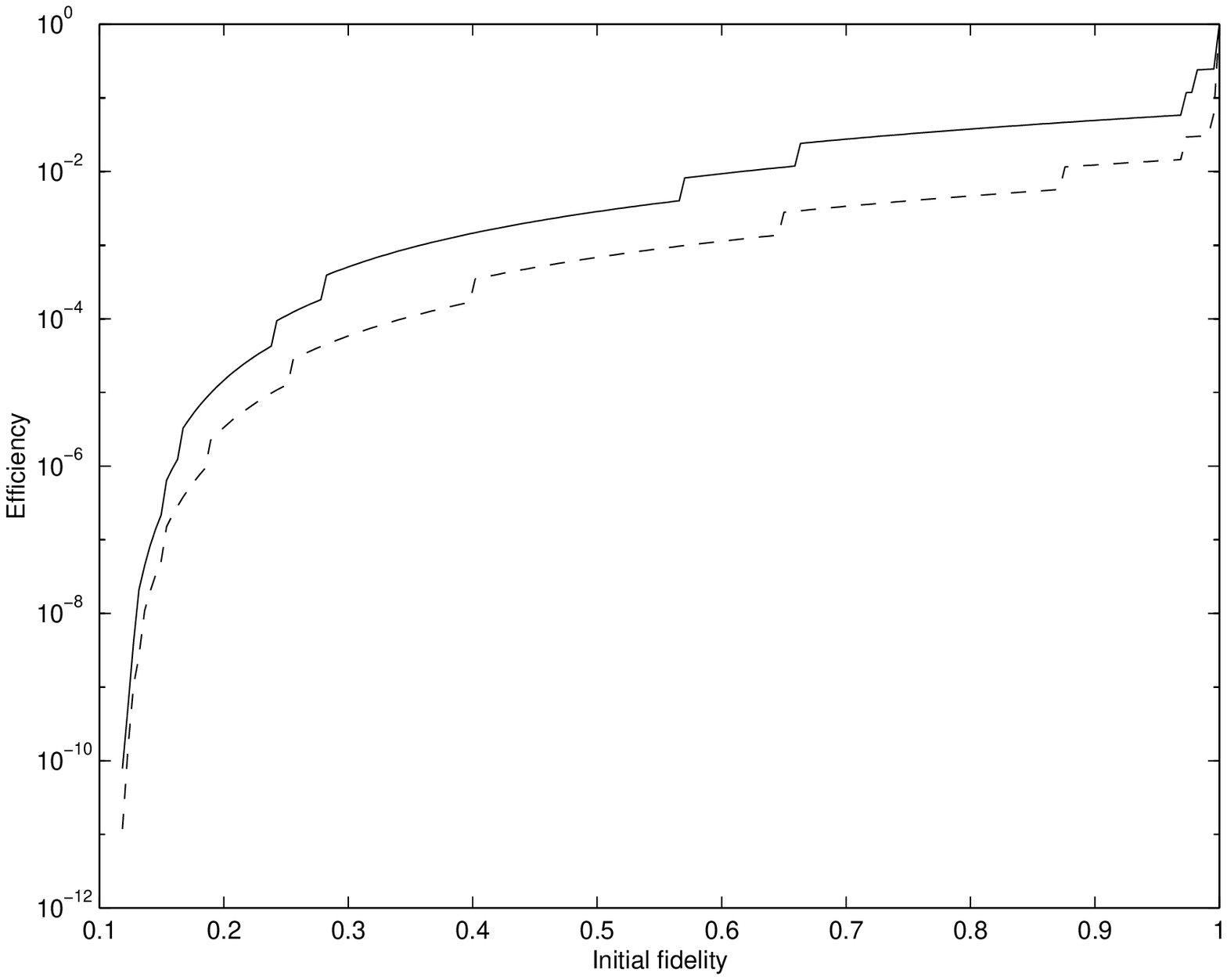,width=8.0cm,clip=}}
\caption{Dependence of the efficiency $\eta$ (compare with Eq. (\ref{eta}))
on the initial 
fidelity $F$ for a fixed final fidelity $F_{final}=1-10^{-5}$ and for
dimension $D = 6$.  The solid line
gives the results of the proposed method and the dashed line the results of
Horodecki's protocol.} \label{Fig.4} 
\end{figure}

\begin{figure}
\centerline{\psfig{figure=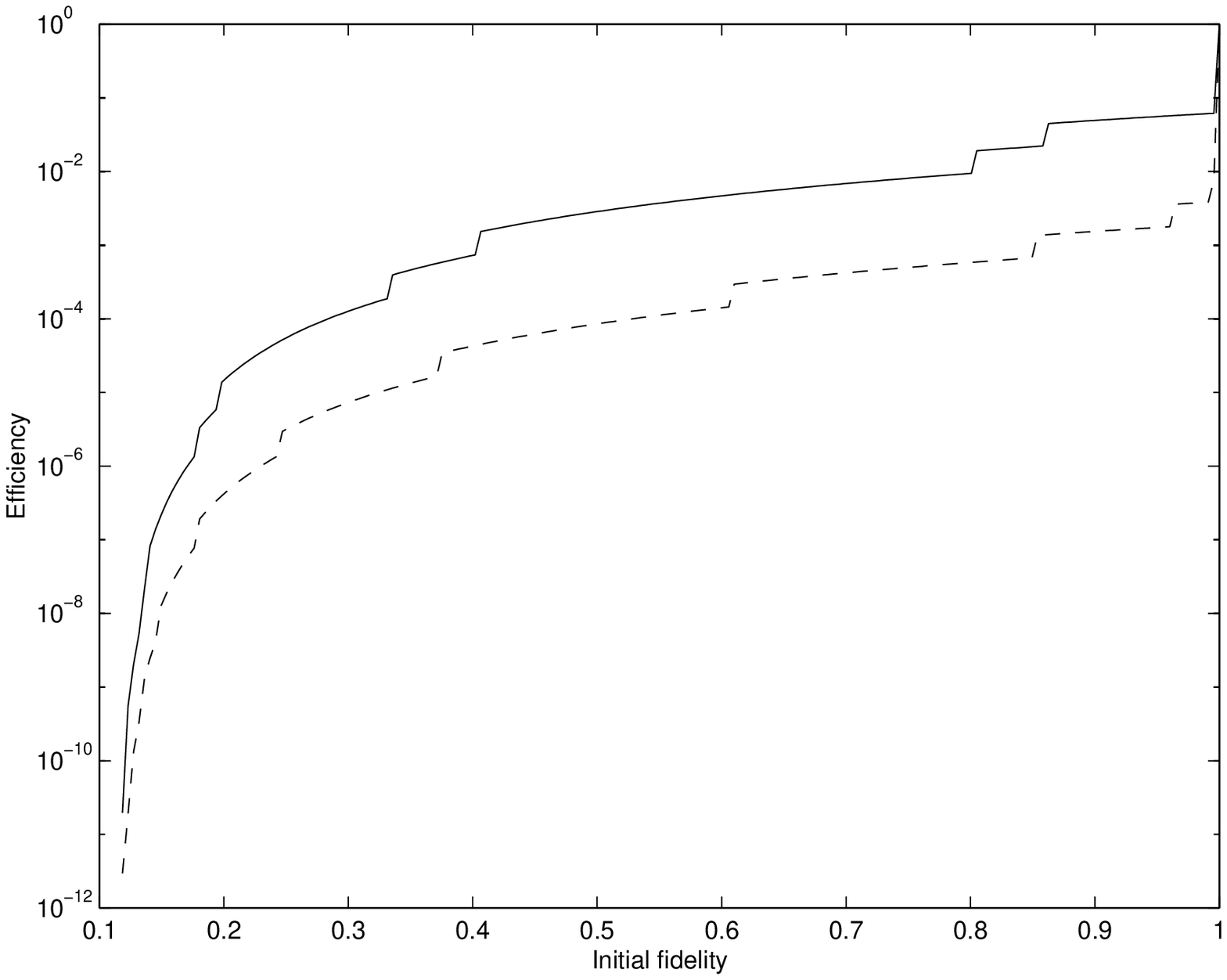,width=8.0cm,clip=}}
\caption{Dependence of the efficiency $\eta$ (compare with Eq. (\ref{eta}))
on the initial 
fidelity $F$ for a fixed final fidelity $F_{final}=1-10^{-7}$ and for
dimension $D = 9$.  The solid line
gives the results of the proposed method and the dashed line the results of
Horodecki's protocol.} \label{Fig.5} 
\end{figure}

\end{document}